# Lateral heterojunctions within monolayer semiconductors

Chunming Huang[1,#,*], Sanfeng Wu[1,#,*], Ana M. Sanchez[2,#,*], Jonathan J. P. Peters[2], Richard Beanland[2], Jason S. Ross[3], Pasqual Rivera[1], Wang Yao[4], David H. Cobden[1], Xiaodong Xu[1,3]

[1]Department of Physics, University of Washington, Seattle, Washington 98195, USA
[2]Department of Physics, University of Warwick, Coventry, CV4 7AL, UK
[3]Department of Material Science and Engineering, University of Washington, Seattle, Washington 98195, USA
[4]Department of Physics and Center of Theoretical and Computational Physics, University of Hong Kong, Hong Kong, China
[#]These authors contribute equally to this work.
[*]Email: SW (swu02@uw.edu); CH (chunmh@uw.edu); AS (A.M.Sanchez@warwick.ac.uk)

**Abstract**

Heterojunctions between three-dimensional (3D) semiconductors with different bandgaps are the basis of modern light-emitting diodes[1], diode lasers[2], and high-speed transistors[3]. Creating analogous heterojunctions between different two-dimensional (2D) semiconductors would enable band engineering within the 2D plane[4-6] and open up new realms in materials science, device physics and engineering. Here we demonstrate that seamless high-quality in-plane heterojunctions can be grown between the 2D monolayer semiconductors $MoSe_2$ and $WSe_2$. The junctions, grown by lateral hetero-epitaxy using physical vapor transport[7], are visible in an optical microscope and show enhanced photoluminescence. Atomically resolved transmission electron microscopy reveals that their structure is an undistorted honeycomb lattice in which substitution of one transition metal by another occurs across the interface. The growth of such lateral junctions will allow new device functionalities, such as in-plane transistors and diodes, to be integrated within a single atomically thin layer.

**Text**

Our lateral heterojunctions are formed within monolayers of the transition metal dichalcogenides[8-10] having formula $MX_2$ (M=Mo, W; X=S, Se, or Te), a class of 2D semiconductors with unique photonic and optoelectronic properties that have great potential for making new kinds of transistors[11], sensors[12], photodetectors[13], LEDs[14], and spin-valleytronic devices[15]. We are able to perform lateral hetero-epitaxy with $MX_2$s because they have the same honeycomb structure[15], in which a planar triangular lattice of M atoms is sandwiched between two planes of X atoms, and because their lattice constants are closely matched (for instance, the mismatch between $WSe_2$ and $MoSe_2$ is only an order of magnitude less than that between graphene and boron nitride). On the other hand, the electronic properties of different monolayer $MX_2$s[8], such as their band gaps[15], work functions, spin-orbit coupling strengths[15,16], and excitonic spectra[9,10,17-19], vary widely, offering a variety of useful heterojunction properties.

We demonstrate lateral hetero-epitaxy using the combination of $WSe_2$ (lattice constant 3.280 Å, optical band gap 1.653 eV) and $MoSe_2$ (3.288 Å, 1.550 eV)[20], both of which have recently been shown to have excellent direct-gap semiconducting and optical properties. We employ direct physical vapor transport[7, 21] using a mixture of $WSe_2$ and $MoSe_2$ powders as the source, typically heated to 950 °C under a 400 sccm flow of hydrogen carrier gas in a one-inch tube furnace (Methods). Monolayers are formed by the vaporized source material crystallizing on a clean silica/silicon substrate, placed downstream.

On inspection in an optical microscope we find that a centimeter-scale patch of the substrate (limited by the furnace geometry and achievable temperature gradient) is covered by either separate monolayer crystals (Fig. 1a) or a semicontinuous monolayer film (Fig. 1b). The separate crystals are often roughly equilateral triangles, up to about ~15 µm in size, interspersed with strips, stars, arrows and other shapes (Supplementary Information S1), similar to those found when growing pure $MX_2$ monolayers[7, 21-24]. Remarkably however, each larger crystal here exhibits two concentric regions with different optical contrast. The width of the paler outer region is similar for all crystals in a given growth, but varies between growths. The continuous films have a speckled appearance (Fig. 1b) which results from merging of the outer paler regions.

In the triangular crystals the darker triangular centers tend to have nearly straight edges, while the outer edges are more jagged. This can be seen at higher resolution in scanning electron microscope (SEM) images such as Figs. 1c and d. Raman spectra (Fig. 1e) collected from the inner and outer regions (indicated by the red and blue dots in Fig. 1d) are characteristic of monolayer $MoSe_2$ and $WSe_2$, respectively[25] (See also Supplementary Information S2 for comparison with exfoliated monolayers). Atomic force microscopy studies also confirm their monolayer nature (Supplementary Information S3). A bright-field transmission electron microscope (BF TEM) image of one of the triangles is shown in Fig. 1f. Its diffraction pattern (inset) is a single six-fold spot pattern implying a single $MX_2$ monolayer crystal lattice ($D_{3h}$ point group symmetry). These observations clearly imply that each crystal is a 2D $MoSe_2$-$WSe_2$ lateral heterostructure formed in a single honeycomb lattice monolayer. This interpretation is confirmed by atomic-resolution TEM studies which will be described below.

We infer that the lateral heterostructures are created in the sequence indicated in Fig. 1g. Early during the growth process $MoSe_2$ is favored and pure monolayer $MoSe_2$ crystals form, but after some time conditions, in particular the vapor pressure of metal atoms, shift in favor of $WSe_2$ growth. The similarity of the two materials permits the epitaxial growth of $WSe_2$ directly on existing $MoSe_2$ crystal edges. Separate $WSe_2$ crystals may also nucleate (the smaller multi-pointed crystals in Figs. 1a-d). The shift in conditions can be explained by the difference in volatility of the source powders combined with depletion of their surfaces. The $MoSe_2$ evaporates more rapidly at first, so initially the metal component of the vapor above the substrate is predominantly Mo. After some time, however, the $MoSe_2$ source granules are depleted of Se on their surface, their volatility is reduced, and the supply of Mo is cut. Meanwhile the supply of W from the $WSe_2$ granules continues or increases as the source warms further, until it is later also cut by surface depletion. The resulting variation of the vapor pressures of Mo and W with time is indicated in Fig. 1h. We emphasize that this process requires no external action during the growth, maintaining ideal conditions for epitaxy with minimal complication. Interestingly, the reproducibility of our results suggests that a more sophisticated setup with independent control of the vapor components could be developed if necessary to create heterojunction sequences programmably for complex device applications.

These new kind of interfaces, which are one-dimensional (1D) in the sense that they are the boundaries of strictly 2D semiconductors, are well suited to precise electron microscopy studies (Methods). Fig. 2a is an SEM image of a continuous film transferred onto a TEM grid. White lines can be seen at the $WSe_2$ boundaries between the crystallites (Supplementary Information S4). Fig. 2b is a low magnification BF TEM image of the film in the vicinity of a nitride-membrane-covered hole in a grid. Here the white lines can be seen to be gaps in the monolayer, and some dendritic darker patches of multi-layer material are also present. The $MoSe_2$ regions appear slightly brighter because only electrons scattered at low angles are detected and due to its lower atomic number, Mo scatters more at low angles than W. Fig. 2c shows a closer look at one of these triangles using annular dark field (ADF) scanning TEM (STEM). Now the inside of the triangle is darker because only electrons scattered to relatively high angles, where W scatters more strongly, are detected[24,26-28]. X-ray energy dispersive spectra (EDS) obtained in the regions defined by the colored polygons are shown in Fig. 2d. The intensities of the Mo, W, and Se peaks, at 2.35, 1.8 and 1.4 keV respectively, are in excellent agreement with $MoSe_2$ in region 1 (green trace) and $WSe_2$ in region 2 (black) as expected. Additionally, in region 3 (brown), the peak heights are equal to the sum of those in regions 1 and 2, implying that the dendritic feature in that region is a $WSe_2$ monolayer growing on top of the $MoSe_2$ monolayer, forming a bilayer $MoSe_2$-$WSe_2$ vertical heterostructure (Supplementary Information S4, S5). We note that the details of the interface vary between growths. For example in Fig. 2c a narrow paler $WSe_2$ band (~30 nm wide) can be seen on the $MoSe_2$ side, indicated by the white arrow. This implies that future mastery of the growth will allow the formation of structures such as 1D quantum wells within the 2D semiconductor.

The 1D interface can be studied with atomic precision using aberration-corrected ADF STEM. Fig. 3a is a high-resolution image taken at the very edge of a dark triangle (also in Supplementary Information S6 and S7). We did not observe any structural instabilities or phase transitions as reported recently in $MoS_2$[28]. It is immediately clear that, remarkably, all the atoms lie on a single $MX_2$ honeycomb lattice. There are no dislocations or grain boundaries, and a Fourier transform of the entire image (Fig. 3b) shows a single six-fold spot pattern implying negligible distortion of the lattice near the interface, i.e., perfect lateral epitaxy.

The intensity contrast is sufficient to identify atoms on the triangular sublattice of the honeycomb containing transition metals. A histogram of the intensities at these sites (Fig. 3c) has two separate peaks, which we plot using a color-scale where the center of the Mo peak is red and that of the W peak is blue. A histogram of the intensities at the chalcogen sublattice sites is superimposed in green. This peak has a low-intensity tail, indicating that some chalcogen sites (Supplementary Information S8) are occupied by only one Se atom[27, 29], possibly due in part to the transfer process and any electron beam damage in the TEM. The intensities at the metal sites in a selected region are plotted in color-scale in Fig. 3d. Fig. 3e shows the intensity along a particular line normal to the interface passing through both M and X sublattices (corresponding to the white dashed line in Fig. 3d), illustrating how the three kinds of atom can be readily distinguished. Fig. 3f is a plot of the intensity over a small rectangular area of the interface using a 3D color-scale chosen to best visualize the distribution of all three atomic species. Fig. 3g shows the inferred identities of the atoms in the same area.

It is clear from these plots that, as is also usually the case with 3D heterojunctions[30], the interface has a finite width. It has the local appearance of a binary alloy, $W_xMo_{1-x}Se_2$, with a steep composition gradient such that $x$ rises from 0 to close to 1 on a scale of several lattice constants. The density of W atoms on the $MoSe_2$ side vanishes within a few lattice constants of the interface, while the density of Mo atoms on the $WSe_2$ side decays more slowly. The sharpness and asymmetry of the interface both suggest that the $MoSe_2$ stops growing before the $WSe_2$ starts and there is subsequently some surface segregation of the Mo during the $WSe_2$ growth. If when a new row of $WSe_2$ is added a fraction $R$ of the Mo atoms from the previous edge row transfer to it, we expect $1 - x = R^n$ for the $n$'th row from the interface[30]. This fits the profile of $x$ well (Fig. 3h), with best fit $R = 0.92$. There is no evidence for clustering of like atoms. Far from the interface on the $WSe_2$ side most Mo substitutions are isolated, such as the one shown in Fig. 3i which corresponds to the small square in Fig. 3a. There are also scattered Se vacancies (Supplementary Information S9) showing no correlation with the interface.

The importance of heterojunctions stems from the substantial difference in electronic and optical properties between the two joined semiconductors. As a test of whether these are preserved, we performed scanning micro-photoluminescence (PL) measurements at room temperature (Methods). Figs. 4a-c are spectrally integrated PL intensity maps for lateral heterostructure crystals with various shapes (more in Supplementary Information S10), while Fig. 4d shows PL spectra taken at the three locations indicated by colored arrows in Fig. 4a. The spectra from the $MoSe_2$ (red) and $WSe_2$ (green) regions are dominated by exciton peaks at 1.57 eV and 1.63 eV respectively, identical to those from the bulk of homogeneous monolayers (Supplementary Information S2, S11). This demonstrates the high optical quality of the material. The spectrum from the heterojunction (black trace) is broader and intermediate in energy, typical for an $MX_2$ ternary compound[31]. Interestingly, the emission from the heterojunctions is brighter than that from the bulk, possibly due to trapping of excitons by defects or enhanced radiative recombination at the interface. Determination of the electrical transport properties for all monolayer semiconductors, including these lateral heterostructures, must wait until reliable contact technologies are developed.

In summary, we have demonstrated the possibility to obtain 2D semiconducting lateral heterostructures with very high crystalline quality through lateral epitaxy. More sophisticated lateral heterostructures, such as 1D quantum wells between 2D barriers, could be made by programmable control of the vapor composition as a function of time during growth. Interesting device physics may also be expected from these atomically thin heterojunctions. For example, spin and valley transport across or along the interface is connected with the unique coupled spin-valley physics in monolayer $MX_2$s. In addition, type II (staggered) band alignment[16] is anticipated at the heterojunction, which may lead to excitons with electrons and holes localized on opposite sides of the interface[32], enabling efficient optoelectronics.

## Methods

**Vapor-solid growth of lateral $WSe_2$-$MoSe_2$ heterostructures**

Typically, a mixture of similar amounts of $WSe_2$ (0.060 g, Alfa Aesar, 99.8 % purity) and $MoSe_2$ powder (0.050g, Alfa Aesar, 99.9 % purity) is placed in an alumina boat in the center of a horizontal quartz tube furnace (Lindburg/blue M) with one-inch tube diameter. After being cleaned with acetone, isopropyl alcohol, and deionized water, the substrate (300 nm $SiO_2$/Si) is placed downstream in a cooler zone (giving ~650-750 °C along the chip during growth). After pumping the tube to about $10^{-3}$ Torr, we heat the furnace to 950 °C. High-purity hydrogen is then passed as a gas carrier at a rate of 400 sccm for 3-10 min. During this growth time, the system pressure is maintained at about 7 Torr. The furnace is finally left to cool to room temperature. For more details of the setup, see Ref. 7 in the main text.

**Wet transfer process**

We use poly(methyl methacrylate) (PMMA)-based transfer techniques to transfer the cyrstals from the growth substrate chips. The chips are spin-coated with PMMA at 3000 r.p.m. for 60 seconds and then submerged in 1M KOH solution. Once the heterostucture-coated PMMA floats on the solution surface, it is transferred to deionized water several times, and then scooped onto a substrate ($SiO_2$ or TEM grid) and dried. The PMMA is removed by baking the sample at 300 $^o$C under atmospheric pressure of $Ar/H_2$ (95%:5%) flowing for 3 h.

**TEM and STEM measurements**

The structure and morphology of the $MoSe_2/WSe_2$ layers were analyzed using a JEOL 2100 and a JEOL ARM200F TEM with $LaB_6$ and Schottky emitters respectively. Experiments were performed with the 2100 operating at 200 kV. STEM analysis on the ARM200F, with probe and image aberration CEOS correctors, was performed at both 80 and 200 kV. ADF STEM images were obtained using a JEOL annular field detector using a fine-imaging probe, at a probe current of approximately 23 pA with a convergence semi-angle of ~25 mrad and an inner angle of 45-50 mrad. To reduce the effects of scan distortions and readout noise, stacks of rapidly-acquired high-resolution STEM images were recorded using the 'StackBuilder' Plug-in for Digital Micrograph by Bernhard Schaffer (http://dmscript.tavernmaker.de) to create high-quality images after drift-corrected summing. The scanning rate of each frame was typically 6 μs per pixel and each frame consists of 512x512 pixels. EDX analyses were performed with probe currents of approximately 200 pA and collected with an Oxford Instruments X-max SDD detector with an area of 80 $mm^2$.

**Photoluminescence (PL) measurements**

We perform room temperature PL measurements using a micro-PL setup, which uses a 40X objective (Olympus) equipped with a XY scanning stage. A 532 nm laser beam (30 μW) is focused onto the sample with about 1 μm spot size. The PL spectrum is then collected by the same objective and detected by a CCD detector (ANDOR) after a grating spectrometer.

**Acknowledgments**

This work is supported by DoE, BES, Materials Science and Engineering Division through XX (DE-SC0008145) and DC (DE-SC0002197). SW is supported partially by the State of Washington through the University of Washington Clean Energy Institute. WY is supported by the Research Grant Council of Hong Kong (HKU705513P), the University Grant Committee (AoE/P-04/08) of the government of Hong Kong, and the Croucher Foundation under the Croucher Innovation Award. XX is grateful for the support of the Research Corporation through a Cottrell Scholar Award. AMS thanks the Science City Research Alliance and the HEFCE Strategic Development Fund for funding support. JJPP acknowledges EPSRC funding through a Doctoral Training Grant.

**Author contributions**

XX conceived the project. CH grew the material. SW led the sample characterization, assisted by CH, JSR and PR, under supervision of XX and DHC. AMS, JJPP and RB performed aberration corrected STEM with samples prepared by SW and CH. SW, DHC, XX and WY wrote the paper with input from all authors.

**Competing financial interests**

The authors declare no competing financial interests.

Figure 1

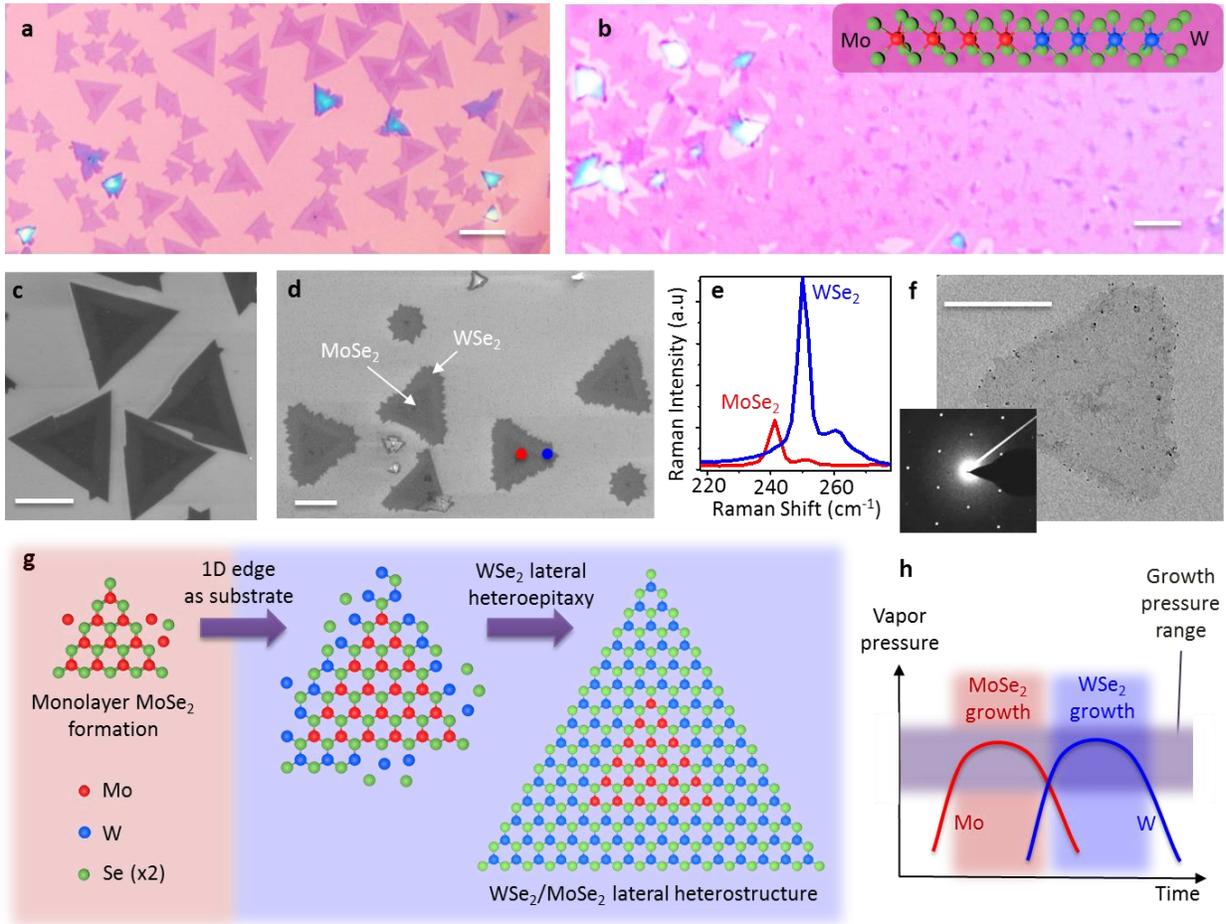

**Figure 1 | In-plane hetero-epitaxy of 2D MoSe$_2$/WSe$_2$ lateral heterostructures. a,** Optical image of triangular heterostructure crystals with concentric regions of different color grown by physical vapor transport using a mixed MoSe$_2$/WSe$_2$ source. **b,** Optical image of a semi-continuous film. Inset: side-view cartoon of an in-plane heterojunction. **c** and **d,** SEM images of heterostructure crystals from two different growths. All scale bars 10 μm. **e,** Raman spectra (514.5 nm laser excitation) taken at the points indicated in **d**. **f,** BF TEM image of an isolated heterostructure. The inset shows its electron diffraction pattern implying a single undistorted lattice. Scale bar: 5 μm. **g,** Schematic illustration of the process of in-plane epitaxial growth of our lateral heteostructures. **h,** Schematic of the atomic vapor pressure variation leading to growth of the two materials in sequence.

Figure 2

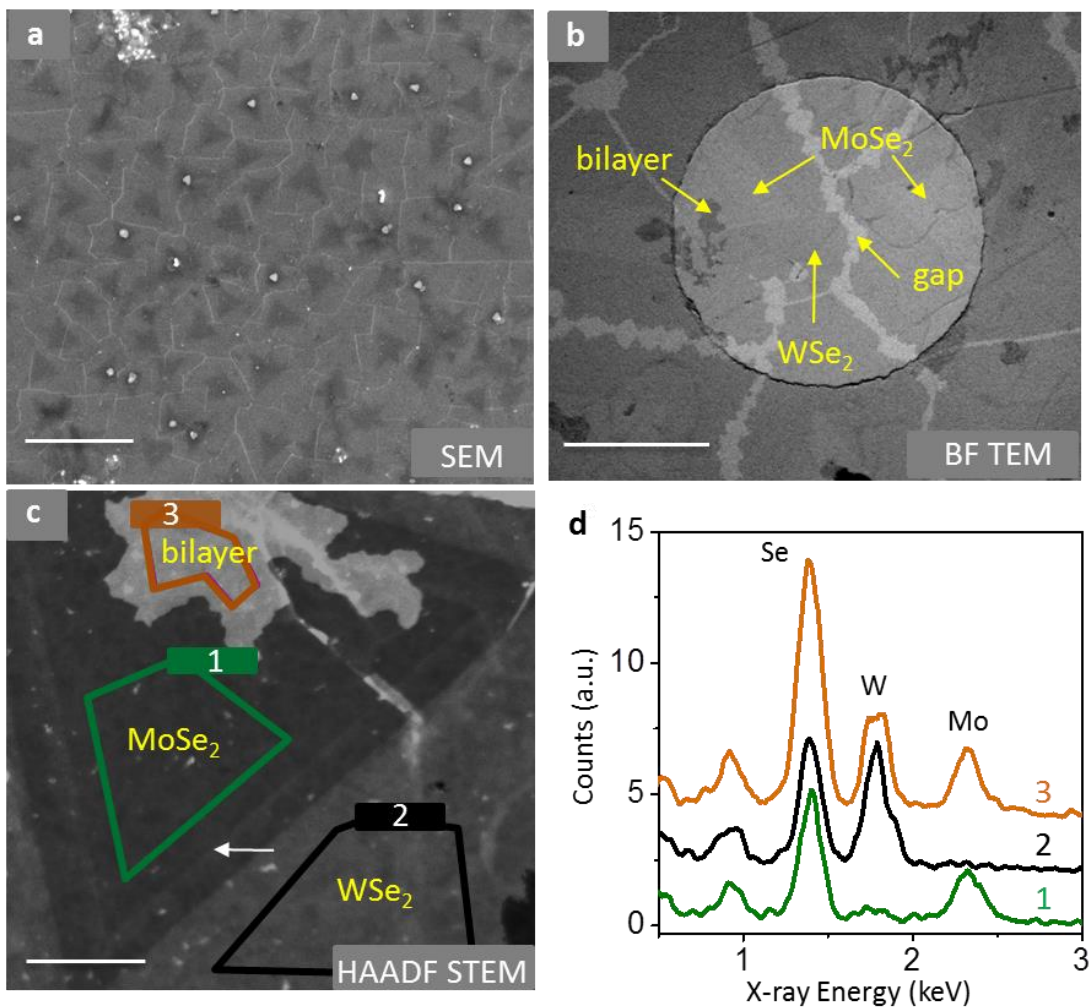

**Figure 2 | Transmission electron microscope (TEM) and energy dispersive spectroscopy (EDS) characterization. a,** SEM image of a film transferred onto a SiO$_2$ substrate. The white lines are gaps at the WSe$_2$ boundaries. Scale bar: 5 μm. **b,** Low-magnification BF TEM image of a film on a TEM grid with 1 μm diameter holes. Faint MoSe$_2$ triangles and gaps at the WSe$_2$ boundaries can be seen. Scale bar: 500 nm. **c,** ADF STEM image of a triangle. The MoSe$_2$ is now darker due to the lower atomic number of Mo than W. The brighter dendritic region is a WSe$_2$ layer growing on top, i.e., a bilayer. Scale bar: 100 nm. **d,** EDS spectra taken form the areas indicated by color-coded polygons in **c**, showing signatures of WSe$_2$ (black), MoSe$_2$ (green) and WSe$_2$/MoSe$_2$ bilayer (brown).

Figure 3

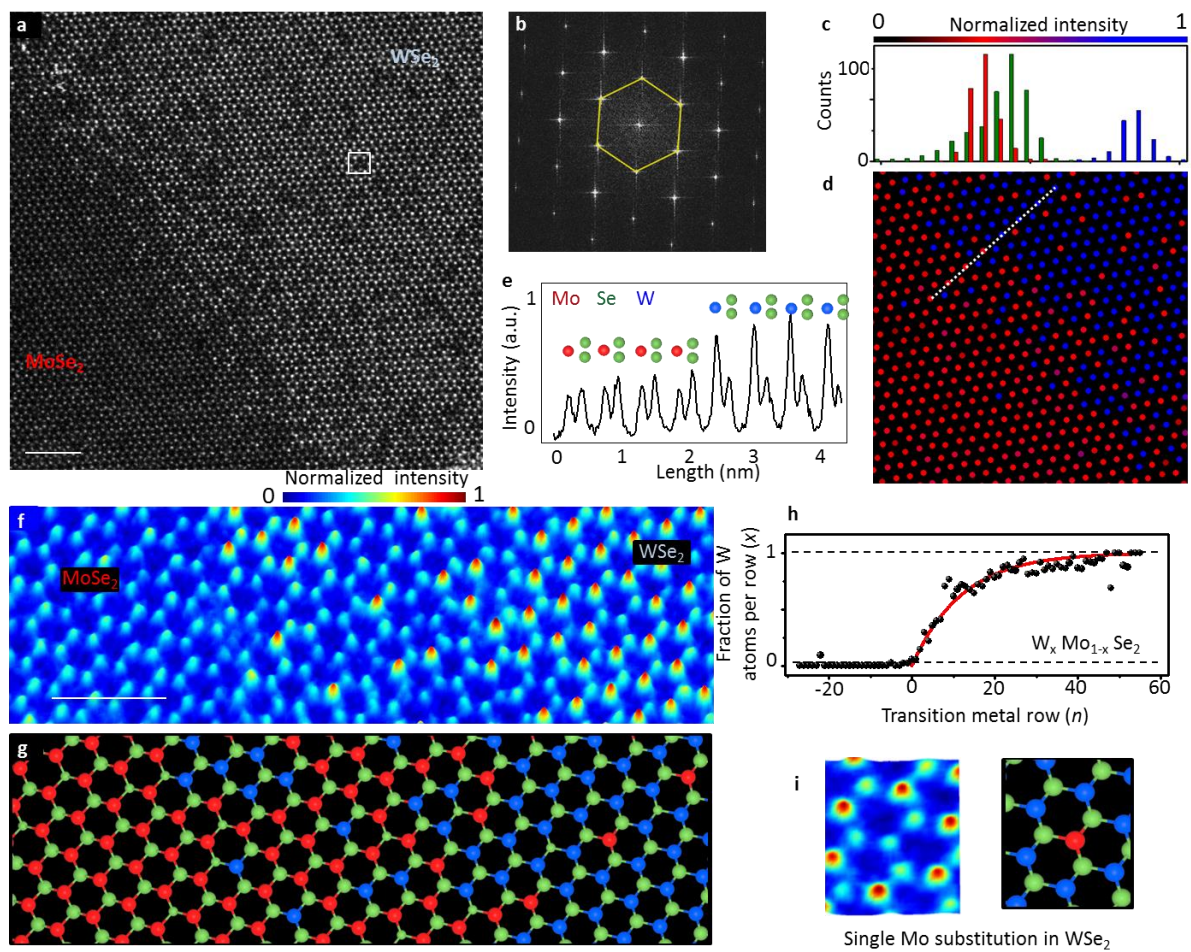

**Figure 3 | 1D interface between 2D semiconductors. a,** High-resolution ADF STEM image of an area of the interface between MoSe$_2$ (darker) and WSe$_2$ (brighter). Scale bar: 2 nm. **b,** Fourier transform of the image in **a** showing a single sharp triangular spot pattern. **c**, Histogram of intensities on the metal sublattice sites (red-blue colorscale) and the chalcogenide sublattice sites (green). **d**, Intensities at the metal sublattice sites in a selected area. **e,** Intensity plot across the interface along the dotted line in **d. f,** 3D color map of a selected area. Scale bar: 1 nm. **g** Atomic species in **f** as identified by scattered electron intensities. **h**, Fraction $x$ of W atoms per row as a function of row index $n$ across the interface, determined from integrated intensities based on the data in **a**. The red solid line is a fit based on a surface segregation model (Ref. 30). **i,** Zoom on a single Mo substitution on the WSe$_2$ side corresponding to the small square in **a**.

Figure 4

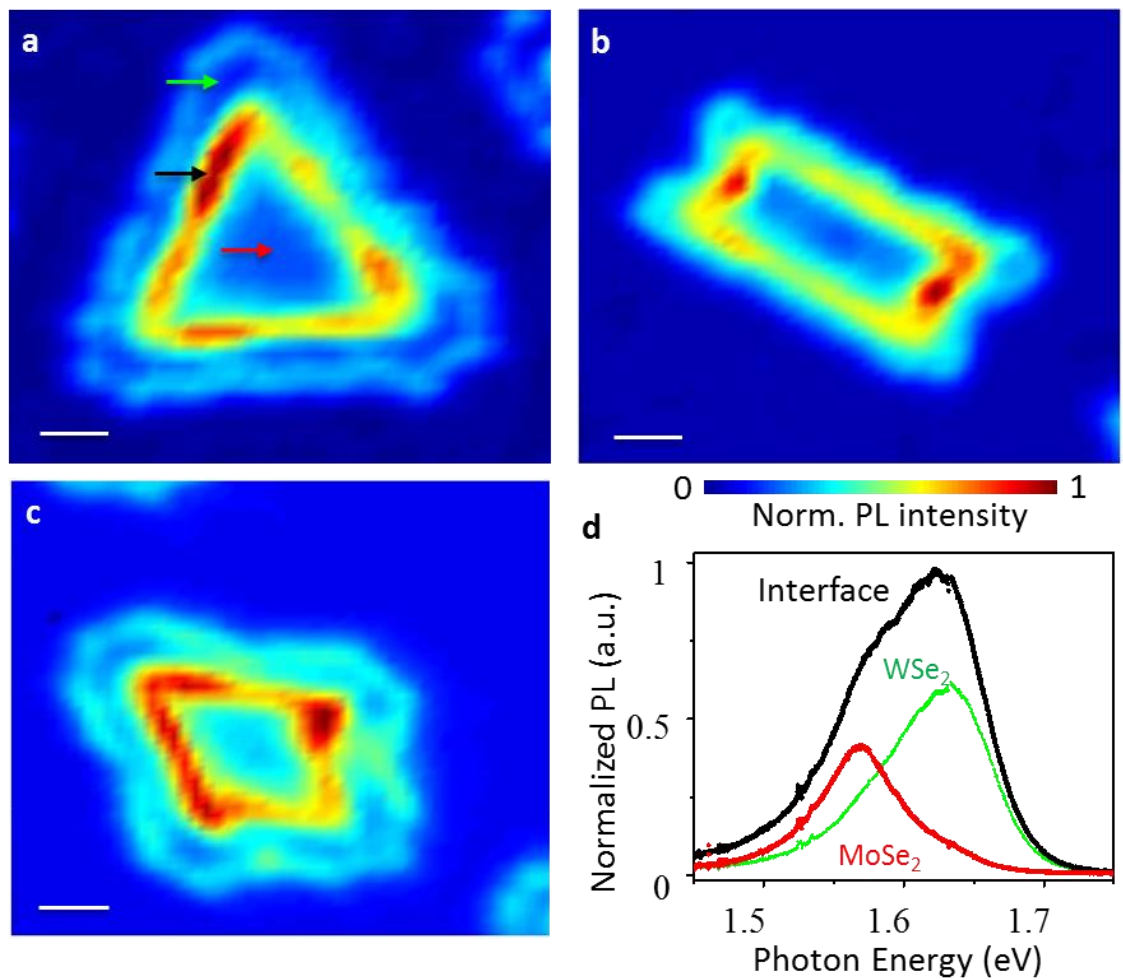

**Figure 4 | Photoluminescence from 1D heterointerfaces.** **a,** 2D PL intensity map of a triangular lateral heterostructure. Intense emission is seen from the 1D interface. Scanning micro-PL was performed with 532 nm laser excitation at room temperature. **b** and **c**, Similar measurements for heterostructures with other shapes. Scale bars: 2 μm. **d,** PL spectra taken at the points indicated in **a**.

# Supplementary Materials

# Lateral heterojunctions within monolayer semiconductors

Chunming Huang[1,#,*], Sanfeng Wu[1,#,*], Ana M. Sanchez[2,#,*], Jonathan J. P. Peters[2], Richard Beanland[2], Jason S. Ross[3], Pasqual Rivera[1], Wang Yao[4], David H. Cobden[1], Xiaodong Xu[1,3]

[#]These authors contributed equally to the work.

*Email: SW (swu02@uw.edu); CH (chunmh@uw.edu); AS (A.M.Sanchez@warwick.ac.uk)

S1. Morphology of the as-grown lateral heterojunctions.

Figure S1a shows a typical distribution of the deposition on a 4.5 cm $SiO_2$/Si chip after vapor-transport growth. We can distinguish four regions on the chip resulting from varying local growth environment along the furnace. While the upstream end has nothing (or tiny dots) deposited, the downstream end can have very thick deposition. In between, triangular crystallites, as well as continuous film, form over about one centimeter. Lateral heterojunctions with different shapes can be found, although most common are roughly equilateral triangles. Figures S1 b-e show SEM images of flakes with different shapes, such as strips, triangles, and bow ties. Figures S1 f-h show optical images of a region with triangles. In many of the shapes we can see concentric regions of different color corresponding to $MoSe_2$ (darker in SEM) and $WSe_2$. Some thicker layers can also be seen, indicating multi-layer deposition.

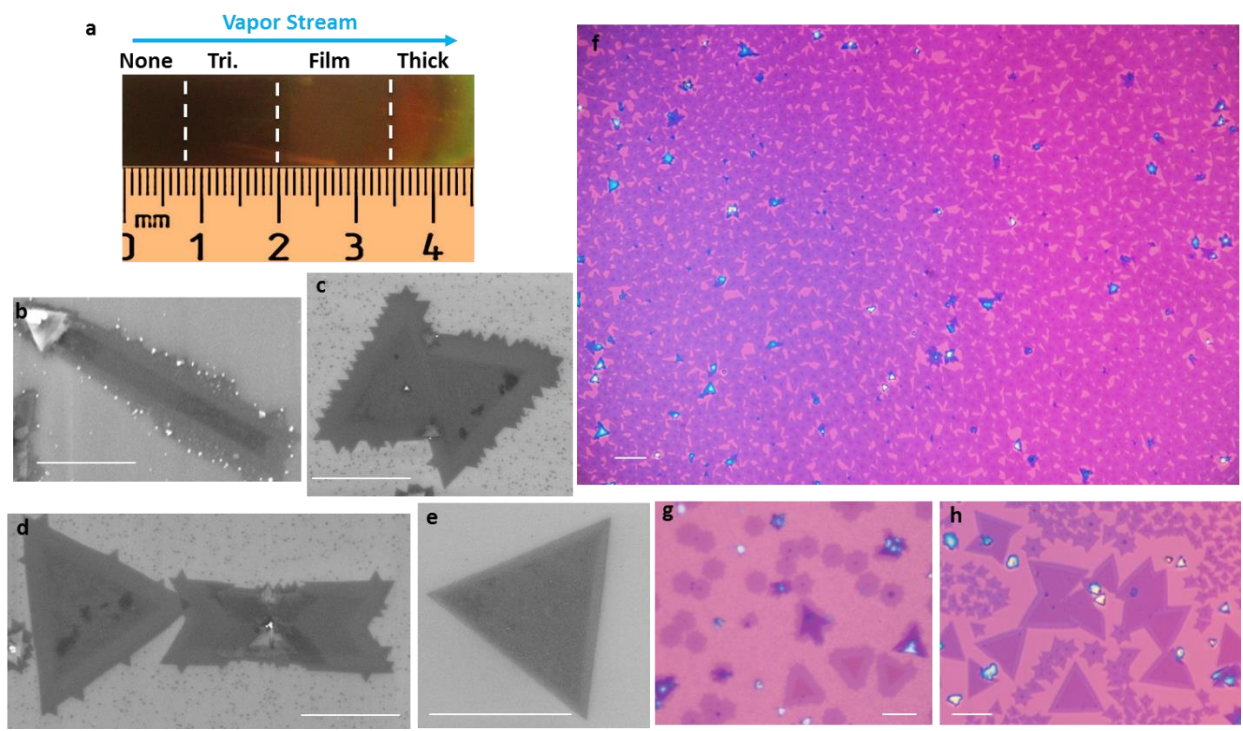

**Figure S1 | Morphology of as-grown lateral heterostructures. a**. Typical overall pattern of deposition on chip. **b-e,** SEM images of monolayer lateral heterojunctions from different growths, and in different morphologies. **f-h,** optical images of a representative growth. All scale bars: 10 μm.

S2. Photoluminescence and Raman scattering from exfoliated monolayer $WSe_2$ and $MoSe_2$

Here we present room temperature PL and Raman data from exfoliated monolayer $WSe_2$ and $MoSe_2$ for comparison with the measurements of grown monolayers in the main text.

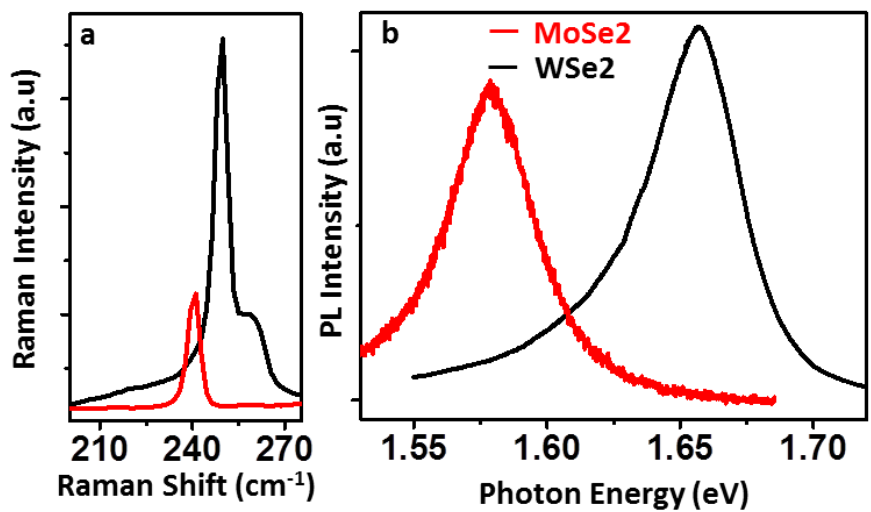

**Figure S2 | a**, Raman, and **b**, PL spectra of exfoliated monolayers. Red: $MoSe_2$. Black: $WSe_2$.

S3. AFM studies of the heterostructures

Figure S3a shows typical AFM images of an as-grown heterostructure, indicating a flat surface over the entire triangle. To perform a thickness measurement, we transfer the as-grown monolayers onto a clean $SiO_2$ substrate, anneal it at 400 $^oC$ for ~2 hours in an $H_2$/Ar environment, and then soak it in a dichloromethane bath (~2 hours), followed by an acetone bath (~1 min) and an IPA bath (~1 min). Figure S3b shows an AFM measurement afterwards. The cracked lines in the $WSe_2$ region appear after annealing. The height of the entire triangle is found to be 0.75 nm. This is further evidence for the monolayer nature of the heterostructures.

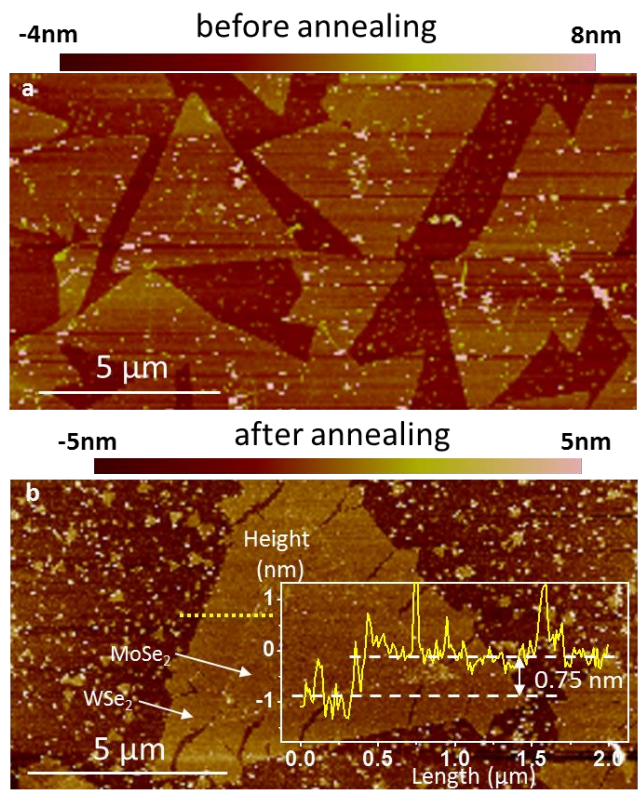

**Figure S3 | AFM measurements of heterostructures. a,** AFM image of the as-grown monolayer heterostructures. **b,** AFM images of an annealed heterostructures. Inset: height measurements along the dashed line showing the atomic monolayer thickness.

S4. TEM characterization of dendritic layers and domains.

Fig. S4a is a typical large-area high-angle ADF STEM image of a film sample. In addition to the darker roughly triangular $MoSe_2$ monolayer regions and the surrounding brighter $WSe_2$ monolayer regions, we also see the following:

(a) Thicker regions in the shape of crystals. EDS measurements show that these are multi-layer $WSe_2$.

(b) Dendritic formations usually associated with the edge of a $MoSe_2$ region. EDS measurements (Fig. 2 in the main text) show that these correspond to a second layer (of $WSe_2$) on top of the first layer (either $MoSe_2$ or $WSe_2$) forming a bilayer. This indicates that in addition to in-plane hetero-epitaxy on the 1D edge it is also possible for a $WSe_2$ layer to initiate and grow on top of an existing layer. In our growths this second layer formation is limited to dendritic shapes that do not cover a large area.

(c) Gaps between $WSe_2$ regions, which are present over the entire film. The polycrystalline nature is confirmed by electron diffraction, such as the pattern in the inset to Fig. S4b collected from the area shown in Fig. S4a. The corresponding bright field (BF) TEM image is shown in Fig. S4b. We see multiple sets of six-fold spot patterns, each of which corresponds to a different crystal orientation in the observed area. By collecting dark field (DF) images from a particular first-order spot we can identify individual crystals. A false-color DF image of an area where four crystals meet is shown in Fig. S4c. We hesitate to call these grains because they do no connect: the DF image also indicates that the gaps between them are essentially empty.

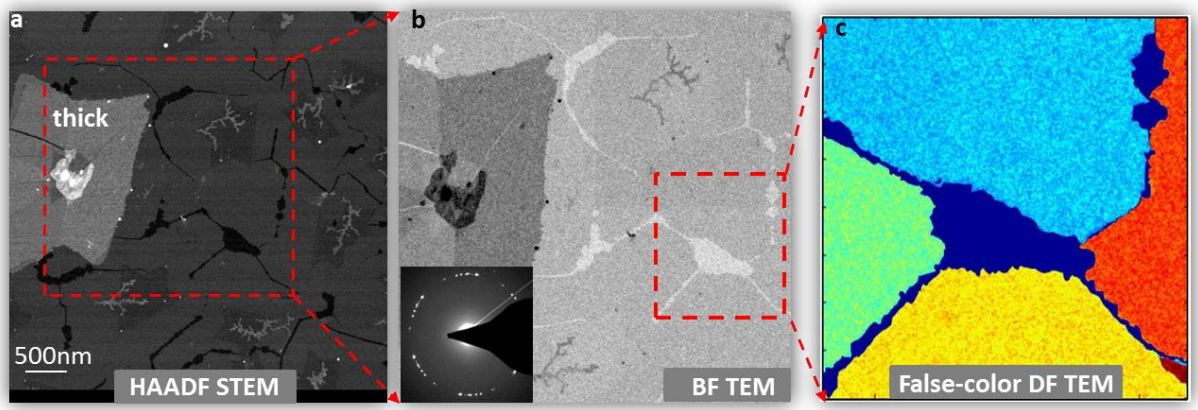

**Figure S4 | TEM characterization of dendritic layers and domains. a**. HAADF STEM image of a selected area where lateral heterostructure, dendritic layers and thick-layer deposition can be seen. **b,** BF TEM image of the rectangular area shown in **a**. The $WSe_2$ and $MoSe_2$ monolayer regions are not easily distinguished, but the dendritic layers are clear. **c,** False-color DF TEM image of the selected area indicated. Colors encode the crystal orientation. Dark blue corresponds to no diffraction peaks, indicating that there are gaps between the crystals.

S5. ADF STEM images of a bilayer MoSe$_2$-WSe$_2$ vertical heterojunction (dendritic structure above).

Fig. S5a shows an atomic-resolution ADF image across the edge of a dentritic bilayer of WSe$_2$ on MoSe$_2$). A line-profile from the selected region (Fig. S5b) indicates passage from a single layer to a bilayer. In the bilayer the upper WSe$_2$ honeycomb appears to perfectly match that of the lower MoSe$_2$ layer. The intensity profile can be explained if the stacking order of this bilayer is C7 type[2], in which the Se atoms of the WSe$_2$ layer are aligned with the Mo atoms of the MoSe$_2$ layer, as illustrated in Fig. S5b.

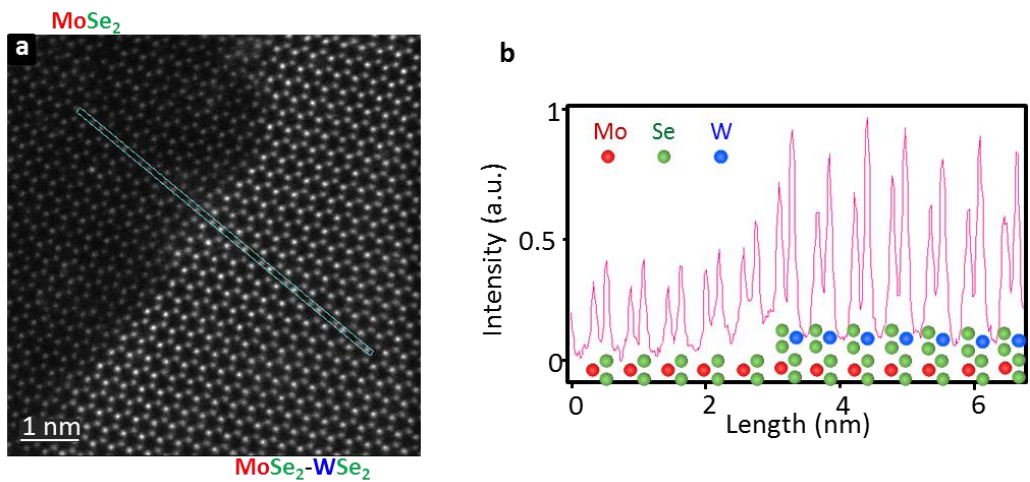

**Figure S5 | Bilayer MoSe$_2$-WSe$_2$ vertical heterostructure. a**. ADF STEM image of the edge of a dendritic layer. **b,** Intensity profile along the line indicated in **a**. The atomic composition is indicated by the cartoon, showing a transition from monolayer MoSe$_2$ to a bilayer MoSe$_2$-WSe$_2$ vertical heterostructure.

S6. Additional ADF STEM images at the 1D interface.

Figure S6a is the same image as Fig.3a in the main text and Fig. 3b is a 3D surface plot of the intensity for this data set. Figs. S6c and d are similar measurements made at another location, demonstrating that all the features of the 1D hetero-interface discussed in the text are common.

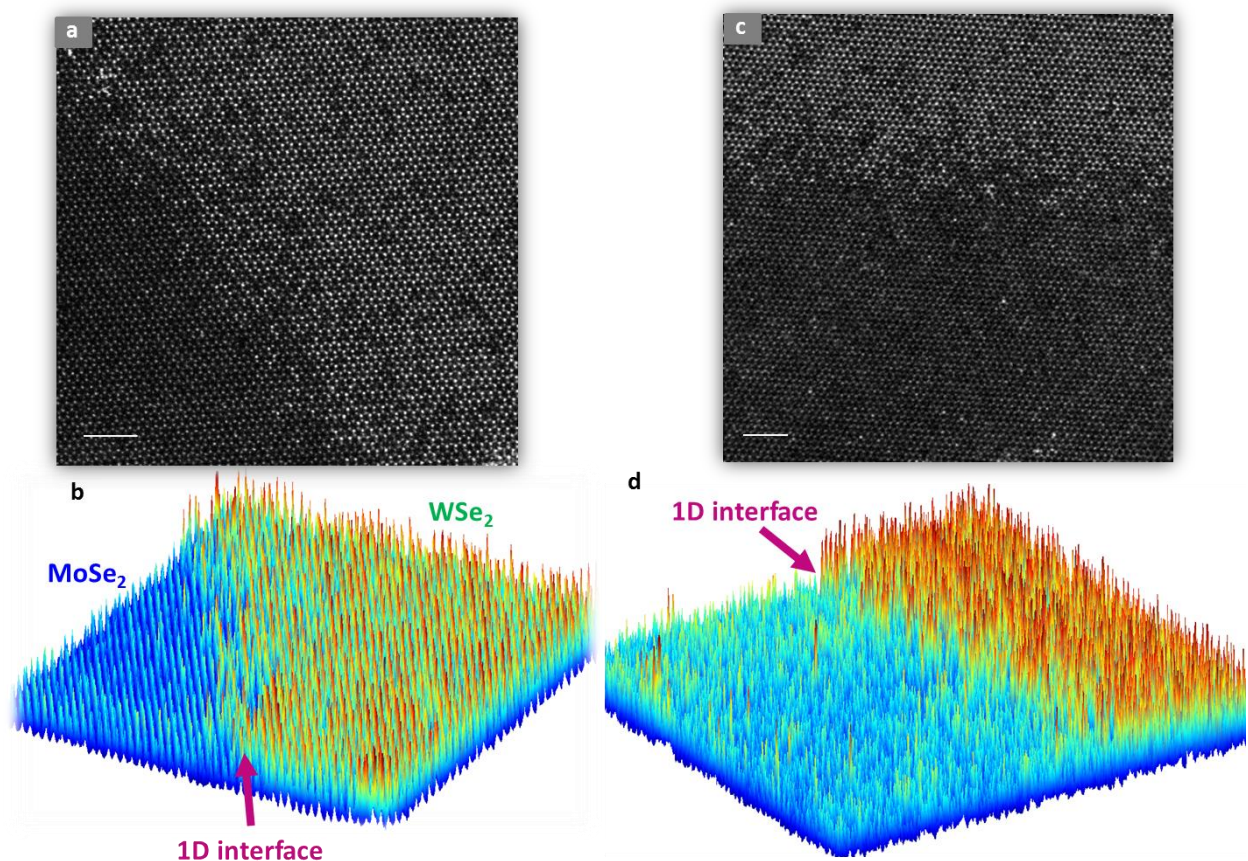

**Figure S6 | a** and **c**, ADF STEM images of the 1D interface in two different places. **b** and **d,** 3D color plots of the same data. Scale bar: 2 nm.

S7. ADF STEM images of WSe$_2$ and MoSe$_2$ monolayer lattice

Figure S7 shows the crystal structures of a, monolayer MoSe$_2$ and b, monolayer WSe$_2$, far from the 1D interface. Both exhibit a honeycomb lattice structure. The transition metal atoms can be easily identified by the intensity in the WSe$_2$ lattice but not in the MoSe$_2$, as expected. The defects visible in the lattice are similar to those reported in MoS$_2$ crystals. We systematically present the defect structure in Section 9 (Fig. S9).

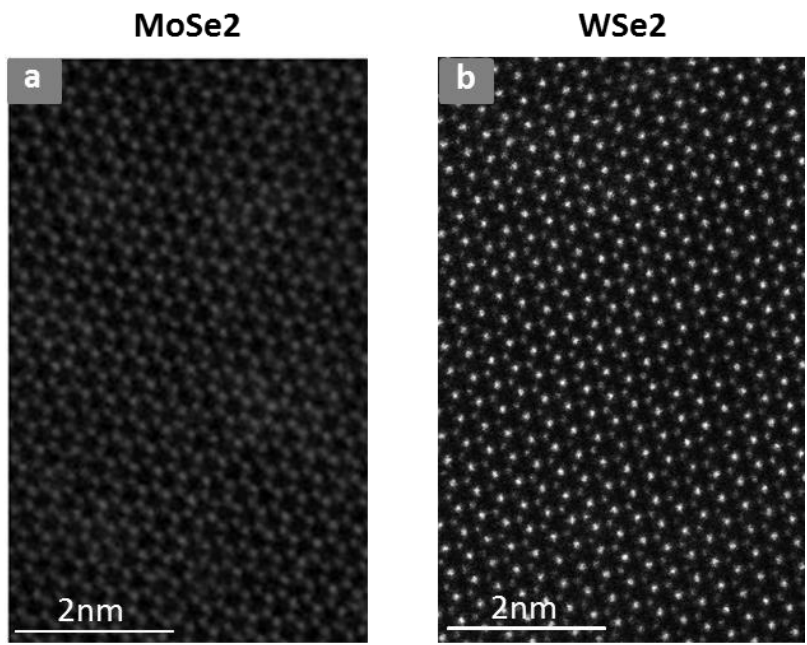

**Figure S7 |** ADF STEM images of **a**, MoSe$_2$, and **b**, WSe$_2$ monolayers.

S8. Atomic statistics at the interface

Here we separate the intensities at the metal and Se sites in the ADF STEM image, using the same data as in as Figs. 3C and D. Fig. S8a is a grayscale image of the measured intensity. Fig. S8b shows the intensities obtained from this data at the centers of the atomic sites on the two triangular sublattices, plotted using a color-intensity scale. This is a combination of Figs. S8c and S8d, where Fig. S8c shows the metal sublattice only (same as Fig. 3D) and Fig. S8d shows the Se sublattice only. The Se sublattice shows a lot of defects (missing one or both atoms at a site) but there is no statistically significant change across the interface.

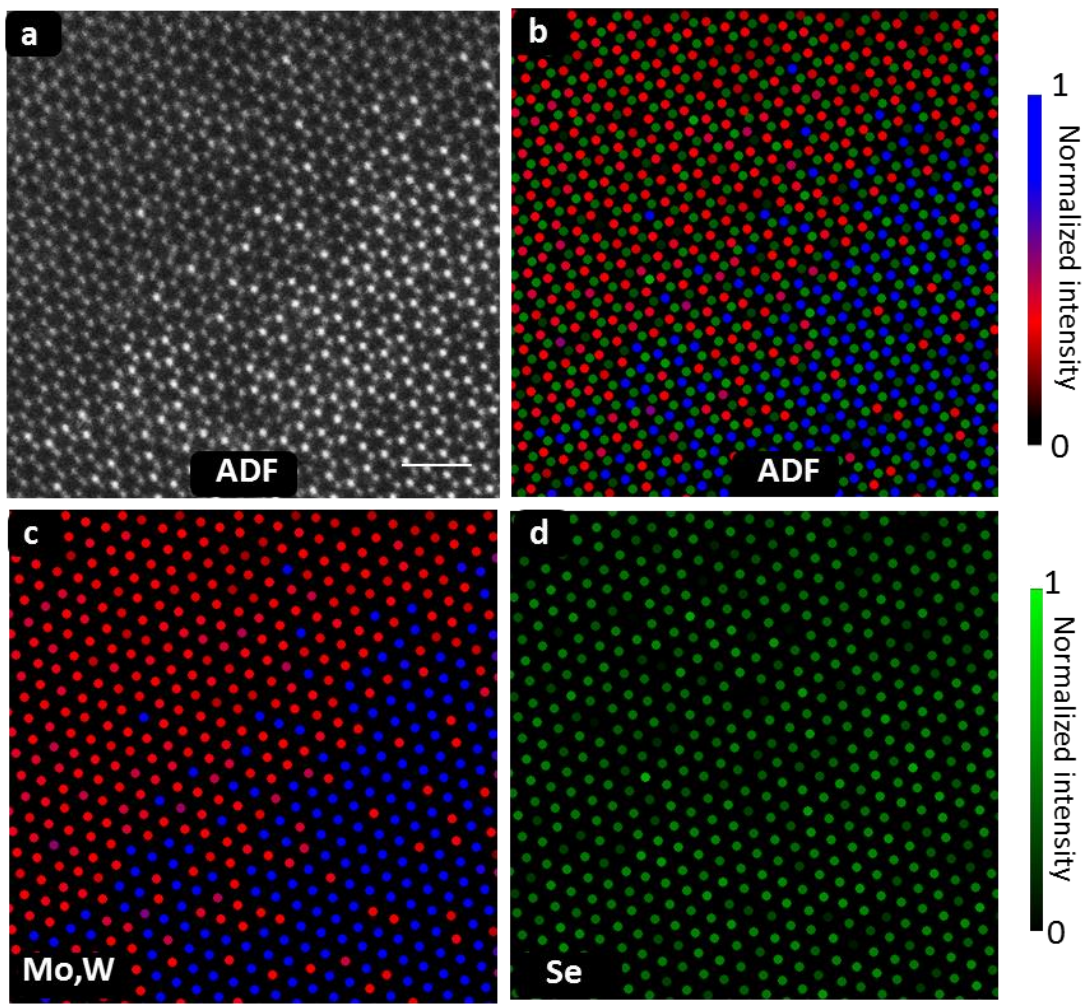

**Figure S8 | Atomic analysis at the interface. a,** ADF STEM image of part of the heterointerface. **b,** Color plot of intensities at the lattice sites derived from **a**. **c,** Intensities for the transition metal (Mo, W) sublattice sites only (same as Fig. 3D). **d,** Intensities at the Se sublattice sites only. Scale bar: 1nm.

S9. More defect configurations (missing metal atoms, missing Se atoms)

In addition to the substitution of the transition metal atoms that defines the interface, some other defects can also be found in both $WSe_2$ and $MoSe_2$ lattice. Those defects are not only present near the interface but also in the bulk of the monolayer flakes (Fig. S7). Here we illustrate representative configurations found near on the $MoSe_2$ side of the interface as indicated in Fig. S9a. Fig. S9b shows a double Se vacancy, where both Se atoms at one honeycomb site are missing. Here two nearby Se sites also have reduced intensity, indicating single missing Se atoms. A single missing Se atom is the dominant defect in our samples. Fig. S9c is an example of such configuration, where two neighboring Se sites are each occupied by only one Se atom. Another occasional defect configuration is a missing metal atom. Fig. S9d shows an example of a missing Mo atom, accompanied by a missing Se atom on each of the two neighboring Se sites.

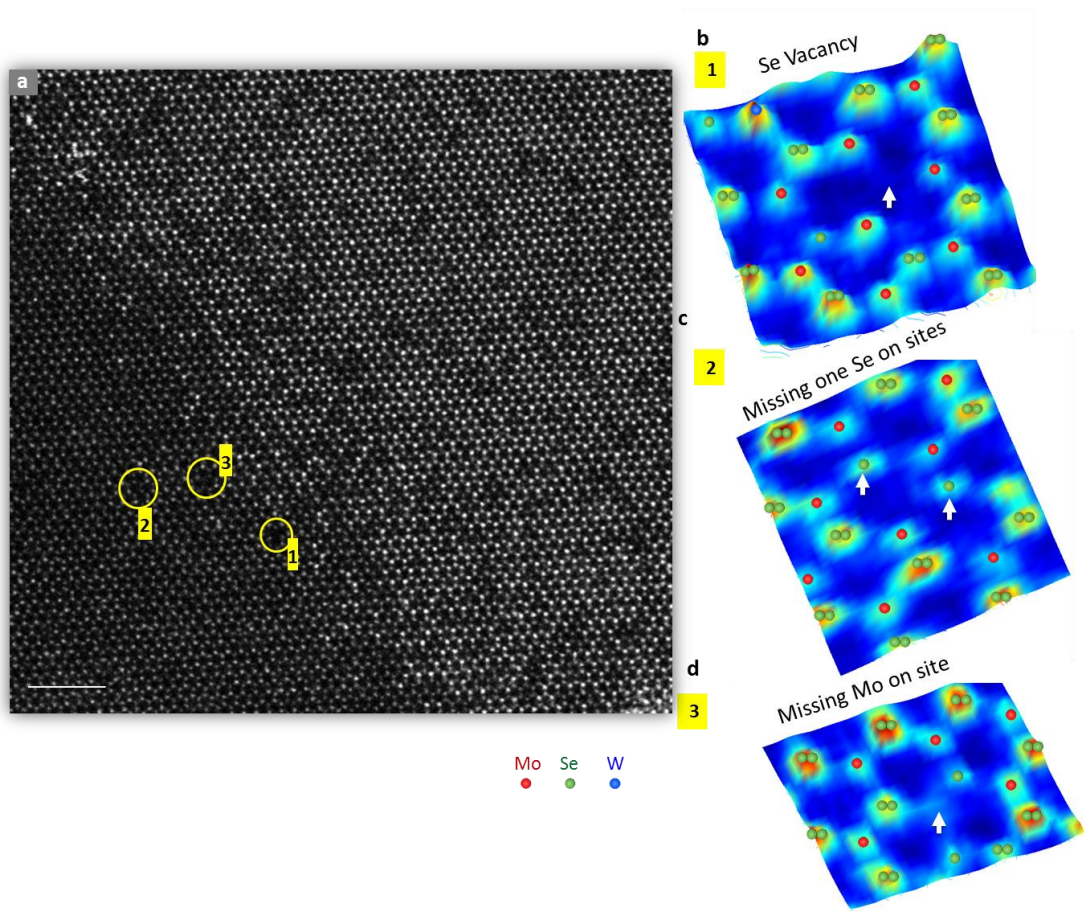

**Figure S9 | Defect configurations. a,** ADF STEM image of the interface shown in Fig. 3A. **b-d,** 3D color plots of several defect configurations, taken from the regions indicated in **a.** Scale bar: 2 nm.

S10. Photoluminescence characterization of additional crystals

PL intensity maps provide an optical visualization of the 1D interfaces, as discussed in the main text. Figure S10a, b, c, and d show more examples with a variety of shapes, including (d) a triangle in "messy" surroundings. All show enhanced PL at the heterointerface.

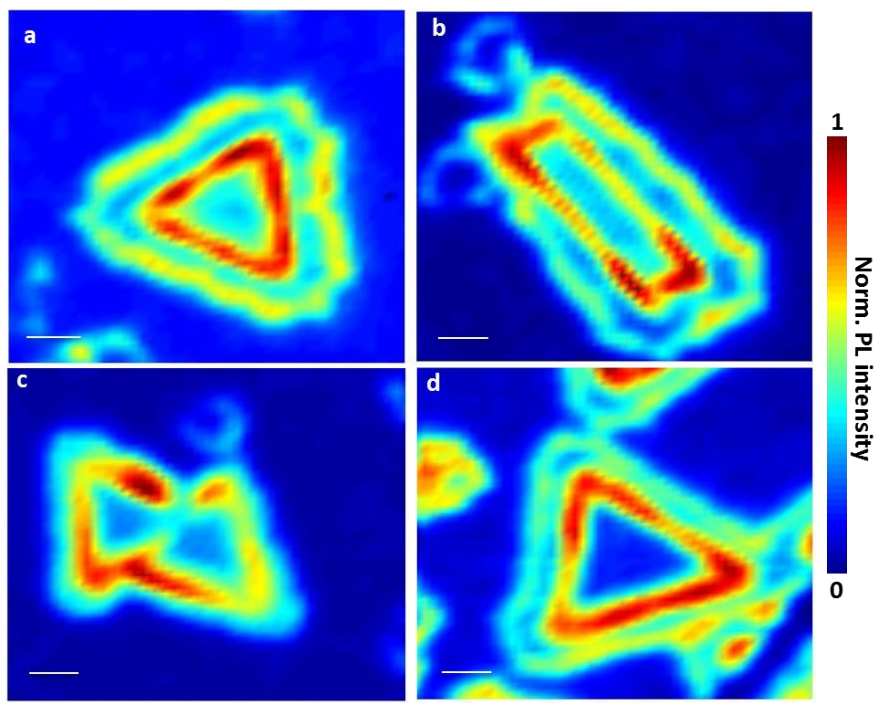

**Figure S10 | Photoluminescence characterization of more flakes. a-d,** Integrated PL intensity maps of monolayer heterostructures with various shapes. All scale bars: 2 μm.

S11. Improved optical quality of the sample after transfer

We find that our as-grown monolayer flakes exhibit shifted PL spectra (at room temperature) compared to exfoliated ones. Such shifts could possibly be due to strain induced during the growth or hybridization with the substrate surface. To remove this effect, we transfer our as-grown flakes to another clean substrate using the technique described in Methods. After transfer the PL peaks are at the usual values, as shown in Fig. S11, for both $WSe_2$ and $MoSe_2$ regions. The $MoSe_2$ emission shifts from 822 nm to 782 nm with almost unchanged intensity, while the $WSe_2$ emission shifts from 783 nm to 758 nm with a greatly enhanced intensity. This suggests that the optical quality of our samples is improved by the transfer process. All the reported PL measurements were conducted after transfer.

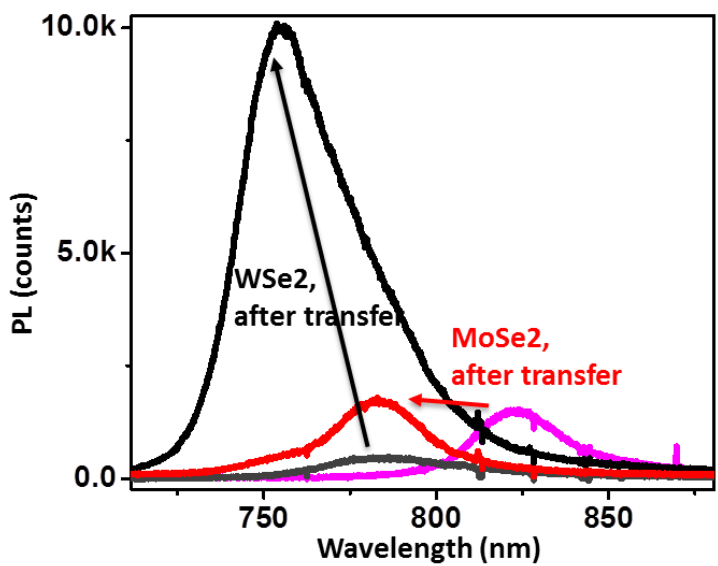

**Figure S11 | Photoluminescence before and after transfer.**

**References**

1, A. M. van der Zande et al., Nat. Mater., 12, 554-561 (2013).
2, K. Kosmider, J. Fernandez-Rossier, Physical Review B, 87(7), (2013).